# A Comparative Agglomerative Hierarchical Clustering Method to Cluster Implemented Course

Rahmat Widia Sembiring, Jasni Mohamad Zain, Abdullah Embong

**Abstract**— There are many clustering methods, such as hierarchical clustering method. Most of the approaches to the clustering of variables encountered in the literature are of hierarchical type. The great majority of hierarchical approaches to the clustering of variables are of agglomerative nature. The agglomerative hierarchical approach to clustering starts with each observation as its own cluster and then continually groups the observations into increasingly larger groups. Higher Learning Institution (HLI) provides training to introduce final-year students to the real working environment. In this research will use Euclidean single linkage and complete linkage. MATLAB and HCE 3.5 software will used to train data and cluster course implemented during industrial training. This study indicates that different method will create a different number of clusters.

—————————— ◆ ——————————

## 1 INTRODUCTION

The main objective of Higher Learning Institutions (HLIs) is to prepare graduates to solve problems by using knowledge and skills, while curriculum must be able to answer the challenge of environment accordance to changes in social needs. Changes can be due to technological advances and global market forces thus the course should be ready to adjust to this reality. Industrial sector also demands the graduates to be able to meet the expectations. Higher Learning Institutions focus to close the year-end students with the industry-related problem, and the most effective way to achieve this goal is by implementing industrial training. Industrial training is a mean to enable students to implement the knowledge and skills learned in relevant field of industries, due this a continuous communication between HLI and industries is important to identify the needs and problems of human resources development.

Industrial training also used to generate and obtain data to evaluate the curriculum of the program, continuous feedbacks from industry and from graduates enable HLIs to develop curriculum to satisfy industrial job needs, it is important to cluster industrial job needs. HLIs must clarify strength of curriculum while offering student to industrial training, so that their student gain success, meanwhile industry can get practical advice for specific problems from HLI.

There are many clustering methods, such as hierarchical clustering method, can further classify into agglomerative hierarchical methods and divisive hierarchical methods [1]. The process of Agglomerative Hierarchical Clustering (AHC) starts with these single observation clusters and progressively combines pairs of clusters, forming smaller numbers of clusters that contain more observations [2]. Then clusters successively merged until the desired cluster structure is obtain [3]. AHC is an elegant technique, to represent the original data set in feature space at multiple levels of abstraction (clustering) because each clustering level results in an abstract representation of the original data set [4].

In this paper, AHC used to cluster frequency of course implemented during industrial training. University find it difficult to determine curriculum is relevant to industry or not, because of limited information and rapid technology growth.

This paper organized into a few sections. Section 2 will present literature review. Section 3 presents methodology, followed by discussion in Section 4 consist of framework for cluster course implemented during industrial training and experiment result, and followed by concluding remarks in Section 5.

————————————————

- *Rahmat Widia Sembiring, is with Faculty of Computer Systems and Software Engineering Universiti Malaysia Pahang, Lebuhraya Tun Razak, 26300, Kuantan, Pahang Darul Makmur, Malaysia, E-mail : rahmatws@yahoo.com*
- *Jasni Mohamad Zain, is Associate Professor at Faculty of Computer Systems and Software Engineering Universiti Malaysia Pahang, Lebuhraya Tun Razak, 26300, Kuantan, Pahang Darul Makmur, Malaysia, E-mail : jasni@ump.edu.my*
- *Abdullah Embong, is Professor at Faculty of Computer Science Universiti Sains Malaysia, 11800 Minden, Penang, Malaysia, E-mail : ae@cs.usm.my*

## 2 Literature Review

Today's rapid technological development has lead to exponential growth of human needs, many aspect of human life increase dependent with technological change, as in the field of medical, financial even up to education. This fact requires HLI has to perform continuous efforts to adapt information from industries, thus graduates meet the needs of industry. Both educators and employers have recognized the importance of educating the professionals that design, develop, and deploy information systems [5].

Therefore, we can use data mining to recognize the relationship between educational institutions and industries, data mining application in education also called the Educational Data Mining. Educational Data Mining (EDM) is the process of converting data from educational systems to useful information that can use by educational software developers, students, teachers, parents, and other educational researchers [6]. Many study discuss EDM, also job opportunities in the industry for college graduates, such as data mining for needy of students based on improved RFM model [6], educating experienced IT professionals by addressing industry's needs [5], jobs for young university graduates [8], measure the relationship between training and job performance [8].

Heuchan and Archibald talk about meeting the needs of the mining engineering sector **Error! Reference source not found.**, extension education to meet market demand [11], training, job satisfaction, and workplace performance [11]. Also development and application of metrics to analyze job data [13], international competitiveness, job creation and job destruction **Error! Reference source not found.**, undergraduate career choice [15], constructing the search for a job in academia [16], students and their performance [16].

This research focuses on industrial training of the final-year students, the objective of industrial training is to experience and understand real life situation in industry and their related environment. Another objective is adequate to practice theory learned in solving real problems, in various aspects such as concept planning, design, construction and administration projects, focus on organizational structure, industrial training and identify the role of certain positions in the industry. Industrial training will perform professionalism and ethics. Clustering can use to explore information about real career world, job requirement for organization structure, business operation and administration functions.

Clustering is concerned with grouping together objects that are similar to each other and dissimilar to the objects belonging to other clusters [18]. Clustered is used to group items that seem to fall naturally together [19]. Various type of clustering: hierarchical (nested) versus partitioned (un-nested), exclusive versus overlapping versus fuzzy, and complete versus partial [20]. Clustering is an unsupervised learning process that partitions data such that similar data items grouped together in sets referred to as clusters, this activity is important for condensing and identifying patterns in data [21].

Clustering techniques apply when there is no class to predict but rather when the instances divide into natural groups. These clusters presumably reflect some mechanism at work in the domain from which instances are drawn a mechanism that causes some instances to bear a stronger resemblance to each other than they do to the remaining instances. Clustering naturally requires different techniques to the classification and association learning methods we have considered so far [19].

Hierarchical clustering creates a hierarchy of clusters, which may represent in a tree structure called a dendogram. The root of the tree consists of a single cluster containing all observations, and the leaves correspond to individual observations. Algorithms for hierarchical clustering are generally either agglomerative, in which one starts at the leaves and successively merges clusters together; or divisive, in which one starts at the root and recursively splits the clusters. Any valid metric use as a measure of similarity between pairs of observations. The choice of which clusters to merge or split is determined by a linkage criterion, which is a function of the pair wise distances between observations [wikipedia].

Agglomerative clustering starts with N clusters, each of which includes exactly one data point. A series of merge operations then followed, that eventually forces all objects into the same group [22]. Hierarchical algorithms find successive clusters using previously established clusters. These algorithms can be either agglomerative ("bottom-up") or divisive ("top-down"). Agglomerative algorithms begin with each element as a separate cluster and merge them into successively larger clusters. Divisive algorithms begin with the whole set and proceed to divide it into successively smaller clusters.

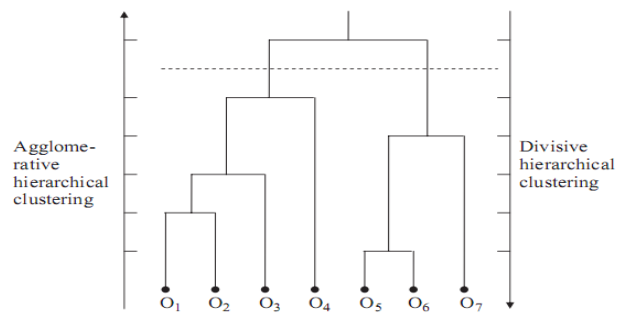

*Figure 1. Example of a dendrogram from hierarchical clustering*

Garcia proposed two specific algorithms obtained from this framework, Hierarchical Compact Algorithm and Hierarchical Star Algorithm areal so introduced. The first creates disjoint hierarchies of clusters, while the second obtains overlapped hierarchies; both algorithms require a unique parameter and the obtained clusters are independent on the data order [22]. A dendrogram describing the relationships between all observations in a data set is useful for understanding the hierarchical relationships in the data, in many situations a discrete number of specific clusters is needed [2].

## 3 METHODOLOGY

The placement of students expected to match a relationship with his/her study program, students expected to implement the knowledge obtained during the study, while industries obtaining information or technological methodologies developed in college. Students need to complete self-assessment questionnaires, distribute during industrial training consisting. Dataset will define like Table-1.

*Table-1 Respondent/student, frequency of course implemented during industrial training*

| Respondent/ Student \ Question/variable | $j_1$ | $j_2$ | .... | $j_n$ |
|---|---|---|---|---|
| $s_1$ | $d_1.. d_n$ | $d_1.. d_n$ | $d_1.. d_n$ | $d_1.. d_n$ |
| ... | $d_1.. d_n$ | $d_1.. d_n$ | $d_1.. d_n$ | $d_1.. d_n$ |
| $s_n$ | $d_1.. d_n$ | $d_1.. d_n$ | $d_1.. d_n$ | $d_1.. d_n$ |

Where (j) is the attribute for the question, (s) are a choice/answer of each question such as implemented of course, knowledge competence and soft skill competence.

## 4 DISCUSSION

### 4.1 Framework for cluster course implemented during industrial training

This research can use to find the cluster of course implemented in industrial training. We proposed a framework, shown in *Figure 2*.

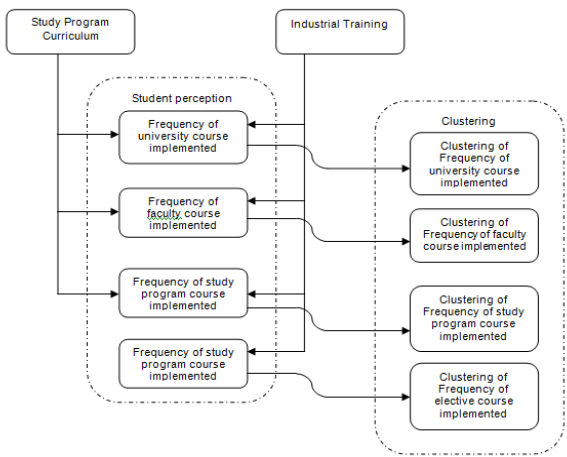

*Figure 2 Framework of Approach*

Training used synthetic data. The data categorized in three parts:
a. Frequency of university course implemented
b. Frequency of faculty course implemented
c. Frequency of study program course implemented
d. Frequency of elective course implemented

MATLAB and HCE 3.5 software use to train data with agglomerative hierarchical clustering, using euclidean distance single linkage compares to complete linkage.

### 4.2 Experiment Result

Agglomerative clustering starts with N clusters, each of which includes exactly one data point. A series of merge operations then followed that eventually forces all objects into the same group. Course of Computer Science Study Program implemented during industrial training, consist of 14 university courses, 17 faculty courses, 9 study program courses, and 3 elective courses.

Using agglomerative hierarchical clustering with euclidean distance single linkage, we had cluster result of the university course implemented as shown in Figure 3.

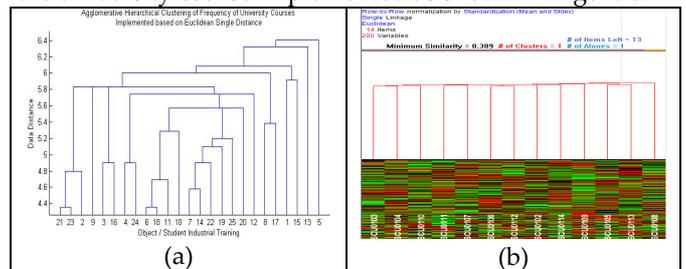

(a)      (b)

*Figure 3 Clustering using Euclidean distance single linkage for Frequency of course implemented with industrial training environment*

From Figure 3(a) use software MATLAB, we can see there are 2 clusters created, while use HCE Figure 3(b) there is 1 cluster with course SCU0108 as a strong course to implement, and SCU0106 and SCU0112 less implemented. Using agglomerative hierarchical clustering with euclidean distance complete linkage, we had cluster result of the university course implemented as shown in Figure 4.

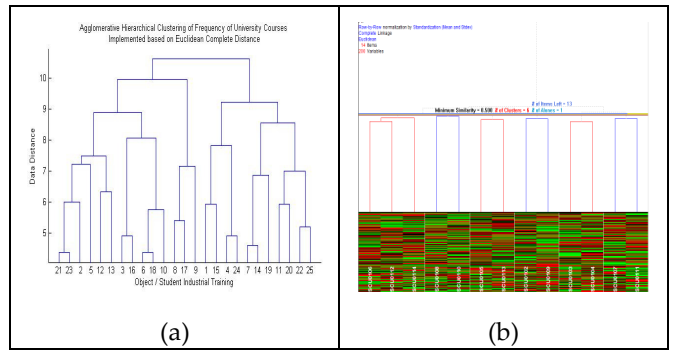

(a)      (b)

*Figure 4 Clustering using Euclidean distance complete linkage for Frequency of university course implemented with industrial training environment*

From Figure 4(a) use software MATLAB, we can see there are 2 clusters created, while use HCE Figure 4(b) there is 6 cluster with course SCU0108 and SCU0110 as a strong courses to implement, and SCU0106 and SCU0112 less implemented. Using agglomerative hierarchical clustering with euclidean distance single linkage, we had cluster result of the faculty course implemented as shown in Figure 5.

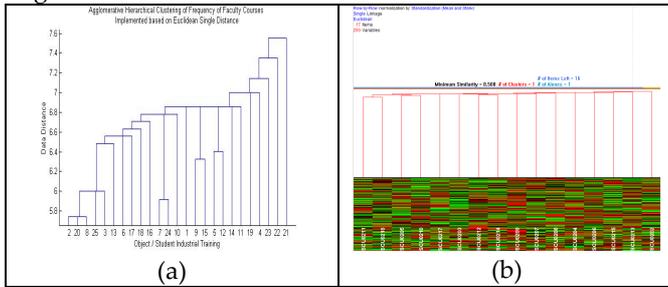

*Figure 5 Clustering using Euclidean distance complete linkage for Frequency of faculty course implemented with industrial training environment*

From Figure 5(a) use software MATLAB, we can see there are 1 clusters created, while use HCE Figure 5(b) there is 1 cluster with course SCU0202 as a strong course to implement, and SCU0211 and SCU0216 are less implemented. Using agglomerative hierarchical clustering with euclidean distance complete linkage, we had cluster result of the faculty course implemented as shown in Figure 6.

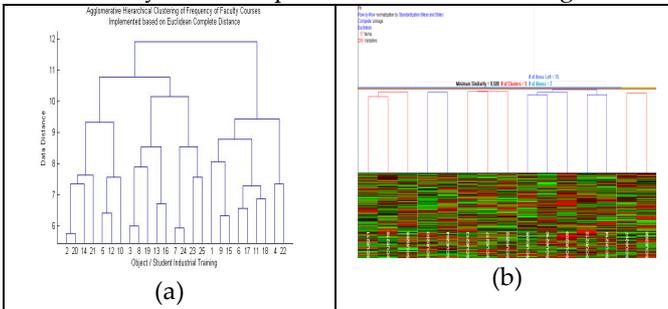

*Figure 6 Clustering using Euclidean distance single linkage for Frequency of faculty course implemented with industrial training environment*

From Figure 6(a) use software MATLAB, we can see there is 1 clusters created, while uses HCE Figure 6(b) there are 5 clusters with course SCU0212 and SCU0214 as a strong course to implement, and SCU0211 and SCU0216 less implemented. Using agglomerative hierarchical clustering with euclidean distance complete linkage, we had cluster result of the study program course implemented as shown in Figure 7.

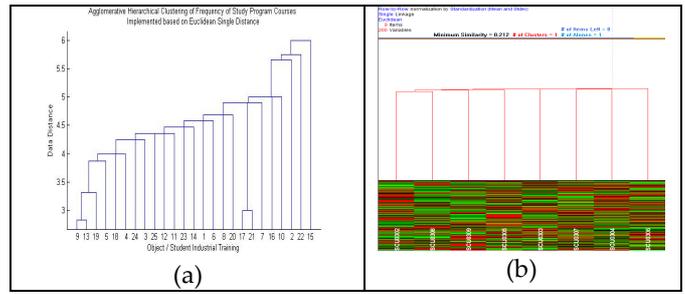

*Figure 7 Clustering using Euclidean distance single linkage for Frequency of study program course implemented with industrial training environment*

From Figure 7(a) use software MATLAB, we can see there are 1 clusters created, while use HCE Figure 7(b) there are 1 cluster with course SCU0306 as a strong course to implement, and SCU0302 and SCU0308 are less implemented. Using agglomerative hierarchical clustering with euclidean distance complete linkage, we had cluster result of the study program course implemented as shown in Figure 8.

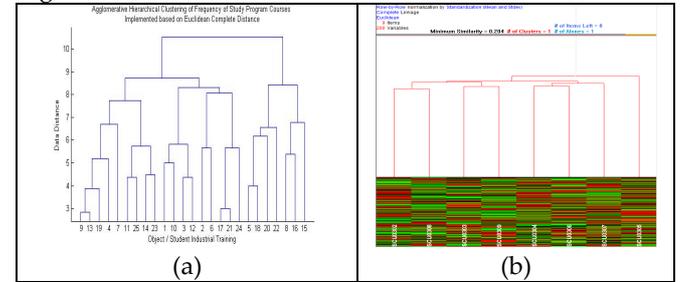

*Figure 8 Clustering using Euclidean distance single linkage for Frequency of study program course implemented with industrial training environment*

From Figure 8(a) use software MATLAB, we can see there are 2 clusters created, while use HCE Figure 8(b) there is 1 cluster with course SCU0305 as a strong course to implement, and SCU0306 and SCU0308 less implemented. Using agglomerative hierarchical clustering with euclidean distance complete linkage, we had cluster result of the elective course implemented as shown in Figure 9.

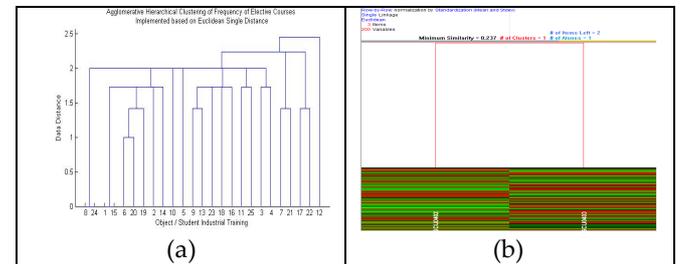

*Figure 9 Clustering using Euclidean distance single linkage for Frequency of elective course implemented with industrial training environment*

From Figure 9(a) use software MATLAB, we can see there is 1 cluster created, while use HCE Figure 9(b) there is 1 cluster with course SCU0402 and SCU0403 as a strong courses to implement, and SCU0401 less implemented. Using agglomerative hierarchical clustering with euclidean distance complete linkage, we had cluster result of the faculty course implemented as shown in Figure 10.

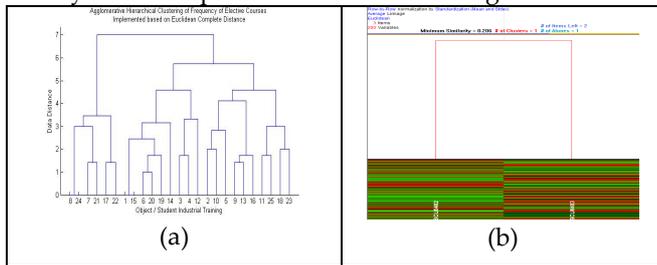

*Figure 10 Clustering using Euclidean distance single linkage for Frequency of course implemented with industrial training environment*

From Figure 10(a) use software MATLAB, we can see there is 1 clusters created, while use HCE Figure 10(b) there are 1 cluster with course SCU0402 and SCU0403 as a strong courses to implement, and SCU0401 is less implemented.

## 5 CONCLUSION AND FUTURE WORK

This study presents a comparative study of the clustering technique Agglomerative Hierarchical using eclidean single linkage and complete linkage for job needs industrial training based on course implemented. This study indicates that different method will create a different number of clusters. Future work includes comparing the performance of different algorithms.

## ACKNOWLEDGMENT

The authors wish to thank Universiti Malaysia Pahang. This work supported in part by a grant GRS090116.

**Rahmat Widia Sembiring** received the degree in Universitas Sumatera Utara in 1989, Indonesia, and master degree in computer science/information technology in 2003 from Universiti Sains Malaysia, Malaysia. He is currently a PhD student in the Faculty of Computer Systems and Software Engineering, Universiti Malaysia Pahang, Malaysia. His research interests include data mining, data clustering, and database.

**Jasni Mohamad Zain** received the BSc.(Hons) in 1989 from in Computer Science (University of Liverpool, England, UK), master degree in 1998 from Hull University, UK, awarded Ph.D in 2005 from Brunel University, West London, UK. She is now as Associate Professor of Universiti Malaysia Pahang, also as Dean of Faculty of Computer System and Software Engineering Universiti Malaysia Pahang.

**Abdullah Embong** received the B.Sc. in 1978 from Universiti Sains Malaysia, Penang, Malaysia, M.S. in 1983 from Indiana University, Bloomington, USA, and awarded Ph.D in 1995 from University of Technology, UK. He is now as Professor of Universiti Sains Malaysia.